\documentclass[preprint]{ptephy_v1}
\usepackage{color}

\newcommand{\ket}[1]{| {#1} \rangle}     
\newcommand{\rbra}[1]{( {#1} |}     
\newcommand{\rket}[1]{| {#1} )}     
\newcommand{\mib}[1]{\boldsymbol {#1}}
\newcommand{\wtilde}[1]{\widetilde{#1}} 

\def\beq{\begin{eqnarray}}
\def\eeq{\end{eqnarray}}
\def\bsub{\begin{subequations}}
\def\esub{\end{subequations}}
\def\b{\begin{equation}}
\def\bs{\begin{split}}
\def\es{\end{split}}
\def\e{\end{equation}}

\begin{document}

\title{A possible trial beyond the conventional random phase approximation for the $su(2)$-Lipkin model 
including the case of non-closed shell system}


\author{Yasuhiko Tsue}
\affil{Deapartment of Mathematics and Physics, Kochi University, Kochi 780-8520, Japan \email{tsue@kochi-u.ac.jp}}

\author{Constan\c{c}a Provid\^encia}

\author{Jo\~ao da Provid\^encia}
\affil{CFisUC, Departamento de F\'{i}sica, Universidade de Coimbra, 3004-516 Coimbra, Portugal}


\author{\hspace{3cm} Masatoshi Yamamura\thanks{These authors contributed equally to this work}}
\affil{Department of Pure and Applied Physics, 
Faculty of Engineering Science, Kansai University, Suita 564-8680, Japan}


\begin{abstract}%
Concerning the $su(2)$-Lipkin model, the calculation of the excitation energy 
to the 1st excited-state gives rise to the following fact: 
The two results based on the exact treatment and the conventional random phase approximation (RPA) 
are in unexpected disagreement. 
In order to remove this discrepancy, the conventional RPA is renewed. 
With the aim of this renewal, the terms, which cannot be dealt with in the conventional 
one, are estimated. 
The basic framework of this new form is not reorganized from that of the conventional one. 
In addition, the $su(2)$-Lipkin model itself is also modified so as to be applicable to the case 
with non-closed shell system. 
To this modification, one more $su(2)$-algebra, which has been already proposed by the present authors, is 
applied. 
Through the use of this algebra, the $su(2)$-Lipkin model 
is applicable to the case with any total fermion number permitted in the model. 
\end{abstract}

\subjectindex{xxxx, xxx}

\maketitle

\section{Introduction}

It is hardly necessary to mention, but the random phase approximation (RPA) has played a central role 
in various fields of many-body theories. 
We quote only two papers for the pioneering works \cite{1}. 
Immediately after the idea of RPA was introduced, theoretical nuclear study changed its nature \cite{2}. 
This method is based on a certain linearization of the equations of motion for the particle-hole pair creation and annihilation 
operators. 
They play a role of the supports in the RPA method. 
In order to define the particle and the hole operators, the free vacuum is indispensable. 
It must be of the closed-shell for a given total fermion number, for example, such as the Hartree-Fock (HF) vacuum. 
Further, for the Hamiltonian widely used, the above-mentioned equations of motion contain 
the terms related to the bilinear forms of the particles and the holes except the supports in RPA. 
In the conventional RPA, these terms are replaced with the expectation values for the free vacuum. 
Through this procedure, we can complete the linearization and we obtain a non-hermitian eigenvalue equation for $\omega$. 
Here, $\omega$ denotes the frequency of the oscillation, which can be regarded as the excitation energy from the ground-state. 
If $\omega^2>0$, this oscillation is stable around the chosen free vacuum and if $\omega^2 <0$, unstable. 
The point $\omega^2=0$ is critical from the stable to the unstable, i.e., usually, 
it is called the phase transition. 
However, the free vacuum commonly used, in many cases, are in non-closed shell. 
In the case where the pairing correlation is strong, we can adopt the BCS-Bogoliubov method. 
In this case, we can formulate the RPA method in the quasi-particle space \cite{3}. 
The bilinear terms for the quasi-particles except the quasi-particle pairs as the supports in RPA are replaced with the 
expectation values for the BCS ground-state. 
The above may be one of the standard understanding of the conventional RPA. 

In relation to the above argument, we must pay attention to the $su(2)$-Lipkin model \cite{4}. 
It supplies a simple model for obtaining the exact results easily such as the excitation energy etc. and, further, 
for checking the validity of the RPA method. 
Since this model obeys the $su(2)$-algebra, it is realized by certain bilinear forms 
in the single-particle fermion operators. 
It consists of two single-particle levels, each of degeneracy $2\Omega$, separated by energy $2\epsilon$. 
Let us take up the following case: 
In the absence of interaction, $2\Omega$ fermion fully occupy the lower level. 
Then, if we regard this state as the free vacuum, the particle and the hole operators can be defined. 
If the interaction is switched on, the fermions can jump from the lower level to the upper one and, 
finally, the $2\Omega$ fermions become to occupy fully the upper level. 
Naturally, the lower becomes vacant. 
The above statement tells us that the free vacuum is nothing but the minimum weight state in the $su(2)$-Lipkin model 
specified by the magnitude of the $su(2)$-spin $s$. 
If the total fermion number is denoted by $N$, the present free vacuum is governed by the condition $N=2\Omega=2s$. 
In this paper, we are interested in the case where the condition $N=2\Omega=2s$ is broken.

%
\begin{figure}[b]
\begin{center}
\includegraphics[height=6.0cm]{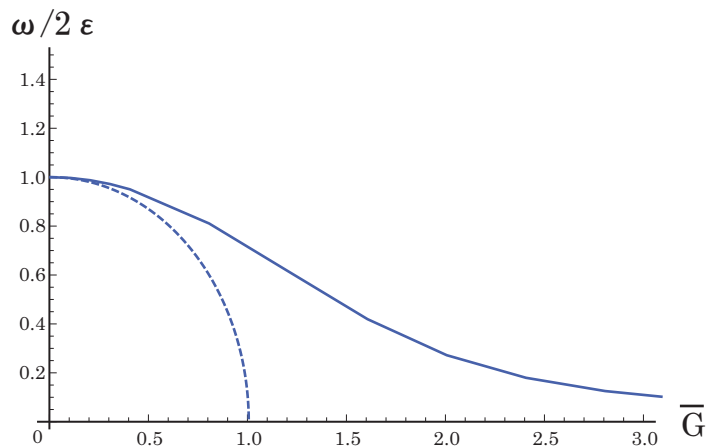}
\caption{The first excited energies from the ground-state energy are depicted as functions 
of $\overline{G}=G/\epsilon\cdot 2\Omega$ for exact (solid curve) and the RPA (dashed curve) results
with $\Omega=5$. 
}
\label{fig:1}
\end{center}
\end{figure}
%

This paper aims mainly at discussing two subjects. 
First is to clarify the relationship among the three quantities $N$, $2\Omega$ and $2s$ appearing in the $su(2)$-Lipkin model. 
For this purpose, we prepare an auxiliary $su(2)$-algebra, three generators of which commute with the three of the original. 
This $su(2)$-algebra has been already taken up for various purposes by the present authors \cite{5}. 
With the aid of this algebra, we can modify the $su(2)$-Lipkin model for treating the case where the condition 
$N=2\Omega=2s$ is broken, i.e., the non-closed shell case. 
It should be noted that one of the generators is related to the total fermion number operator which the 
Lipkin model does not contain. 
Combining these two $su(2)$-algebra, we obtain, for example, the relation $2s\leq N \leq 4\Omega-2s$. 
Therefore, in the case with $s=\Omega$, automatically, we have $N=2\Omega$. 
It suggests us that we can treat the cases with $N<2\Omega$ and $N>2\Omega$. 
The above is the first subject. 
Second is to develop a possible trial beyond the conventional RPA for the Lipkin model. 
Figure 1 depicts the frequency $\omega/(2\epsilon)$ given by the conventional RPA and 
the exactly calculated excitation energy from the ground- to the 1st excited-state as functions of 
$\overline{G}(=G/\epsilon\cdot 2\Omega)$ for $\Omega=5$. 
Here, $G$ denotes the strength of the interaction defined in the relation (\ref{79}). 
We are forced to see that the discrepancy is quite conspicuous and cannot see any phase transition \cite{6}. 
For the above situation, we must develop a new idea. 
In this paper, we will present a possible form for making up the above discrepancy without breaking down 
the basic framework of the conventional RPA. 
For this aim, we focus on the parts replaced with the expectation values given by the free vacuum. 
Our idea is to replace these parts with certain forms obtained through the relation of the 
Casimir invariant. 
With the help of this procedure, we can take up the effects let out in the conventional RPA. 
As an additional remark, we note that the problem discussed in this paper may be old-fashioned, but, 
until the present, any clear-cut solutions for them have been not obtained. 
This is our motivation for the present work. 
Concerning this point, we are interested in the paper\cite{7}.

%
Of course, there are many attempts to extend the RPA itself to include the higher effects such as an anharmonicity 
from before. 
For example, 
in order to extend the HF and RPA methods, the variational approach to the many-body problem was devised by introducing 
an appropriate trial state and the second RPA was derived in a useful form for obtaining approximate solutions\cite{JdP1}. 
Also, by using the Green function formalism, the correction to the RPA was derived as a microscopic theory dealing with 
an anharmonic effects\cite{JdP2}. 
In the words of the Beliaev-Zelevinsky boson expansion, 
the generalized RPA was considered by taking into account the boson-boson correlations\cite{JdP3}. 
Further, it was shown that, if the variational parameters $a$, appearing in the 
generator-coordinate method, $\Psi(x)=\int \phi(x;a)f(a)da$, depends on the time, 
the variational parameters oscillate with the same frequency as that derived by the RPA\cite{JdP4}, 
while the relation between the generator-coordinate method and the RPA was indicated in several authors\cite{Add}. 

In the next section, a modified form of the $su(2)$-Lipkin model is presented. 
Another generators of the $su(2)$-algebra adding to the original ones in the $su(2)$-Lipkin model are introduced. 
In section {\bf 3}, a renewed random phase approximation is formulated and the non-closed shell system can be treated. 
The conventional RPA is included in the renewal of the RPA developed in this section. 
In section {\bf 4}, various numerical results obtained by the renewed RPA are shown. 
Especially, the renewed random phase approximation presents a good result 
for the first-excited energy compared to the conventional RPA. 
The last section is devoted to concluding remarks.  


\section{A modified form of the $su(2)$-Lipkin model}

The $su(2)$-Lipkin model, which may be the simplest many-fermion model, 
consists of two single-particle levels with the degeneracies $2\Omega$. 
Here, $2\Omega$ denotes an integer and we specify the two levels by the index $l$ with 
$l=0$ and 1. 
The single-particle states are specified by an appropriate labeling. 
In this paper, we adopt the labeling $\mu=1,\ 2,\cdots ,\ 2\Omega$ for 
$(0,\mu)$ and $(1,\mu)$. 
The basic operators in this model are as follows: 
\beq\label{1}
{\wtilde S}_0=\frac{1}{2}\sum_{\mu}\left(
{\tilde c}_{1\mu}^*{\tilde c}_{1\mu}-{\tilde c}_{0\mu}^*{\tilde c}_{0\mu}\right)\ , \quad
{\wtilde S}_+=\sum_{\mu}{\tilde c}_{1\mu}^*{\tilde c}_{0\mu}\ , \quad
{\wtilde S}_-=\sum_{\mu}{\tilde c}_{0\mu}^*{\tilde c}_{1\mu}\ .
\eeq
Here, ${\tilde c}_{l\mu}^*$ and ${\tilde c}_{l\mu}$ denote the creation and the annihilation fermion operators 
in the state $(l,\mu)$, respectively. 
Needless to say, the hermitian conjugate are given as 
\beq\label{2}
{\wtilde S}_0^*={\wtilde S}_0\ , \qquad
{\wtilde S}_{\pm}^*={\wtilde S}_{\mp}\ . 
\eeq
The set $({\wtilde S}_{0,\pm})$ obeys the $su(2)$-algebra: 
\beq\label{3}
\left[\ {\wtilde S}_+\ , \ {\wtilde S}_-\ \right]=2{\wtilde S}_0\ , \qquad
\left[\ {\wtilde S}_0\ , \ {\wtilde S}_\pm\ \right]=\pm{\wtilde S}_\pm\ . 
\eeq
The Casimir operator ${\wtilde {\mib S}}^2$ is given in the form 
\beq\label{4}
{\wtilde {\mib S}}^2=\frac{1}{2}\left[\ {\wtilde S}_+\ , \ {\wtilde S}_-\ \right]_++{\wtilde S}_0^2\ .  
\eeq
Naturally,  $({\wtilde S}_{0,\pm})$ commutes with ${\wtilde {\mib S}}^2$.

The Hamiltonian treated in this paper is of the form which has been adopted conventionally:
\beq\label{5}
{\wtilde H}=2\epsilon {\wtilde S}_0-G\left(\left({\wtilde S}_+\right)^2+\left({\wtilde S}_-\right)^2\right)\ . 
\eeq
Here, $2\epsilon$ and $G$ denote the energy difference between the two single-particle levels and the interaction strength, 
respectively. 
We regard the $l=0$ level as the lower than the $l=1$. 
In this paper, we will treat the case with 
\beq\label{6}
\epsilon \geq 0\ , \qquad G \geq 0\ .
\eeq

First task in the algebraic approach for the Hamiltonian (\ref{5}) is to give 
the minimum weight state. 
Without loss of generality, $2s$ fermion can occupy the single-particle states $\mu=1,\ 2,\cdots ,\ 2s$ in the level $l=0$, 
because the single-particle energy does not depend on $\mu$. 
This state, which denotes $\ket{s}$, satisfies the condition 
\beq\label{7}
{\wtilde S}_-\ket{s}=0\ , \qquad 
{\wtilde S}_0\ket{s}=-s\ket{s}\ , \qquad
{\wtilde{\mib S}}^2\ket{s}=s(s+1)\ket{s}\ . 
\eeq
Certainly, the state $\ket{s}$ is a minimum weight state of the present $su(2)$-model. 
Then, we have the state $\ket{s,s_0}$ in the form 
\beq\label{8}
\ket{ss_0}=\left({\wtilde S}_+\right)^{s+s_0}\ket{s}\ , \quad
(s_0=-s,\ -s+1,\cdots ,\ s-1,\ s)
\eeq
Conventionally, this model has been investigated in the case with $2s=N=2\Omega$ 
($N$; total fermion number) under the name of the study of collective motion observed in the closed-shell system. 
With the use of the state $\ket{s}$, we can investigate the case with $N=2s < 2\Omega$. 
Then, the next task is to present the minimum weight state in the condition $N > 2s$. 
If this task realizes, the basic part of the study of the $su(2)$-Lipkin model may be completed.

With the aim of obtaining the minimum weight state in $N > 2s$, we introduce another $su(2)$-algebra 
under the same single-particle level scheme as that in the present model: 
\beq\label{9}
& &{\wtilde \Lambda}_0=\frac{1}{2}\sum_{\mu}\left({\tilde c}_{1\mu}^*{\tilde c}_{1\mu}+{\tilde c}_{0\mu}^*{\tilde c}_{0\mu}\right)
-\Omega\ , \quad
{\wtilde \Lambda}_+=\sum_{\mu}{\tilde c}_{1\mu}^*{\tilde c}_{0\mu}^*\ , \quad
{\wtilde \Lambda}_-=\sum_{\mu}{\tilde c}_{0\mu}{\tilde c}_{1\mu}\ , \nonumber\\
& &\left({\wtilde \Lambda}_0^*={\wtilde \Lambda}_0\ , \quad
{\wtilde \Lambda}_{\pm}^*={\wtilde \Lambda}_{\mp}\right)\ . 
\eeq
The set $({\wtilde \Lambda}_0,{\wtilde \Lambda}_{\pm})$ obeys also the $su(2)$-algebra :
\beq\label{10}
\left[\ {\wtilde \Lambda}_+\ , \ {\wtilde \Lambda}_-\ \right]=2{\wtilde \Lambda}_0\ , \qquad
\left[\ {\wtilde \Lambda}_0\ , \ {\wtilde \Lambda}_\pm\ \right]=\pm {\wtilde \Lambda}_\pm \ .
\eeq
As was already mentioned, the algebra $({\wtilde \Lambda}_{\pm,0})$ has been taken up 
for the various purposes by the present authors \cite{5}.

The Casimir operator ${\wtilde {\mib \Lambda}}^2$ is given as 
\beq\label{11}
{\wtilde {\mib \Lambda}}^2=\frac{1}{2}\left[\ {\wtilde \Lambda}_+\ , \ {\wtilde \Lambda}_-\ \right]_+
+{\wtilde \Lambda}_0^2 \ . 
\eeq
The noticeable relation which controls these two algebras is as follows:
\beq\label{12}
\left[\ {\rm any\ of}\ {\wtilde \Lambda}_{0,\pm}\ , \ {\rm any\ of}\ {\wtilde S}_{0,\pm}\ \right]=0 \ .
\eeq
First, we note that $\ket{s}$ is also a minimum weight state of $({\wtilde \Lambda}_{0,\pm})$:
\beq\label{13}
{\wtilde \Lambda}_-\ket{s}=0\ , \qquad
{\wtilde \Lambda}_0\ket{s}=-\lambda\ket{s}\ .
\eeq
Naturally, $\ket{s}$ is an eigenstate of ${\wtilde {\mib \Lambda}}^2$ with the eigenvalue $\lambda(\lambda+1)$. 
Here, we must notice that $\lambda$ is given as 
\beq\label{14}
\lambda=\Omega-s\ , \quad {\rm i.e.,}\quad 
s+\lambda=\Omega \ .
\eeq
Then, we have $\ket{\lambda\lambda_0}$ in the form 
\beq\label{15}
\ket{s,\lambda\lambda_0;s+\lambda=\Omega}=\left({\wtilde \Lambda}_+\right)^{\lambda+\lambda_0}\ket{s}\ . 
\eeq
The state (\ref{15}) satisfies 
\beq\label{16}
{\wtilde \Lambda}_0\ket{s,\lambda\lambda_0;s+\lambda=\Omega}=\lambda_0\ket{s,\lambda\lambda_0;s+\lambda=\Omega}\ . 
\eeq
With the use of the relation (\ref{12}), we can see that the state (\ref{15}) is the minimum weight state of 
$({\wtilde S}_{0,\pm})$: 
\beq\label{17}
{\wtilde S}_-\ket{s,\lambda\lambda_0;s+\lambda=\Omega}=0\ , \quad
{\wtilde S}_0\ket{s,\lambda\lambda_0;s+\lambda=\Omega}=-s\ket{s,\lambda\lambda_0;s+\lambda=\Omega}\ .
\eeq
The definition of ${\wtilde \Lambda}_0$ shown in the relation (\ref{9}) and its property 
(\ref{16}) give us that the total fermion number $N$ in the state (\ref{15}) is of the form 
\beq\label{18}
N=2\Omega+2\lambda_0 \ .
\eeq
Since $-\lambda \leq \lambda_0 \leq \lambda$, we have 
\beq\label{19}
2\Omega-2\lambda \leq N \leq 2\Omega+2\lambda\ , \quad 
{\rm i.e.,}\quad 
2s \leq N \leq 4\Omega-2s\ . 
\eeq
%
\begin{figure}[b]
\begin{center}
\includegraphics[height=8.0cm]{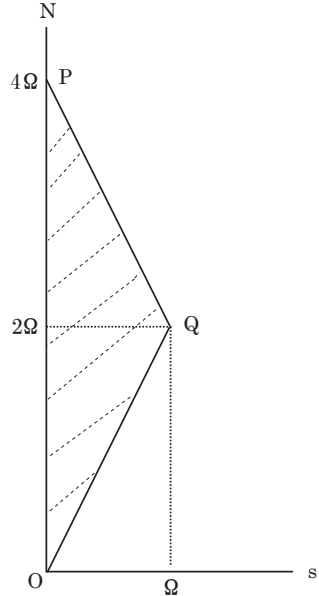}
\caption{The inside area of OPQ denotes an area with $2s\leq N \leq 4\Omega-2s$ in Eq. (\ref{19}). 
}
\label{fig:2}
\end{center}
\end{figure}
%
In such a way, we obtained the minimum weight state in the condition $N > 2s$ and 
the orthogonal set is given as 
\beq\label{20}
\ket{ss_0,\lambda\lambda_0;s+\lambda=\Omega}&=&
\left({\wtilde S}_+\right)^{s+s_0}\left({\wtilde \Lambda}_+\right)^{\lambda+\lambda_0}\ket{s}
\qquad ({\rm in}\ \lambda=\Omega-s) \nonumber\\
&=&\left({\wtilde S}_+\right)^{s+s_0}\left({\wtilde \Lambda}_+\right)^{\frac{1}{2}(N-2s)}\ket{s} \nonumber\\
&=&\left({\wtilde \Lambda}_+\right)^{\frac{1}{2}(N-2s)}\ket{ss_0}\ . 
\eeq
Here, we used the relation (\ref{18}).
If paying attention only to the fact that the set $\{ {\wtilde S}_{\pm,0}\}$ obeys the $su(2)$-algebra, 
we are able to obtain the eigenvalues of the Hamiltonian (\ref{5}) in the space $\{\ket{ss_0}\}$ 
given in the relation (\ref{8}). 
The results are not expressed in terms of $N$. 
In other words, there does not exist the reason why we restrict ourselves to the closed-shell system for the 
applicability of the Lipkin model. 
Any point existing in the interia of the triangle OPQ obeys the $su(2)$-Lipkin model. 
The point Q corresponds to the closed-shell system.

With the aim of making a connection between $({\wtilde S}_{0,\pm},{\wtilde \Lambda}_{0,\pm})$ and the minimum weight state 
$\ket{s}$ more clearly, we reform our algebra. 
First, we decompose ${\wtilde S}_{0,\pm}$ into the following two parts: 
\beq\label{21}
{\wtilde S}_{0,\pm}={\wtilde S}_{0,\pm}^{(1)} +{\wtilde S}_{0,\pm}^{(2)}\ , 
\hspace{9cm}
\eeq
\bsub\label{22}
\beq
& &{\wtilde S}_0^{(1)}=\frac{1}{2}\sum_{\mu}{}^{(1)}({\tilde c}_{1\mu}^*{\tilde c}_{1\mu}-{\tilde c}_{0\mu}^*{\tilde c}_{0\mu}) \ , 
\quad
{\wtilde S}_+^{(1)}=\sum_{\mu}{}^{(1)}{\tilde c}_{1\mu}^*{\tilde c}_{0\mu}\ , \quad
{\wtilde S}_-^{(1)}=\sum_{\mu}{}^{(1)}{\tilde c}_{0\mu}^*{\tilde c}_{1\mu}\ , \quad
\label{22a}\\
& &{\wtilde S}_0^{(2)}=\frac{1}{2}\sum_{\mu}{}^{(2)}({\tilde c}_{1\mu}^*{\tilde c}_{1\mu}-{\tilde c}_{0\mu}^*{\tilde c}_{0\mu}) \ , 
\quad
{\wtilde S}_+^{(2)}=\sum_{\mu}{}^{(2)}{\tilde c}_{1\mu}^*{\tilde c}_{0\mu}\ , \quad
{\wtilde S}_-^{(2)}=\sum_{\mu}{}^{(2)}{\tilde c}_{0\mu}^*{\tilde c}_{1\mu}\ . \quad
\label{22b}
\eeq
\esub
Here, the sums $\sum_{\mu}^{(1)}$ and $\sum_\mu^{(2)}$ are defined as 
\beq\label{23}
\sum_{\mu}{}^{(1)}X_{\mu}=\sum_{\mu=1}^{2s}X_\mu\ , \qquad
\sum_\mu{}^{(2)}X_\mu=\sum_{\mu=2s+1}^{2\Omega}X_\mu\ .
\eeq
They obey 
\beq\label{24}
\left[\ {\rm any\ of}\ {\wtilde S}_{0,\pm}^{(1)}\ , \ {\rm any\ of}\ {\wtilde S}_{0,\pm}^{(2)}\ \right]=0 \ .
\eeq
The operation of ${\wtilde S}_{0,\pm}^{(2)}$ on $\ket{s}$ leads us to 
\beq\label{25}
{\wtilde S}_{0,\pm}^{(2)}\ket{s}=0\ . 
\eeq
Then, we have 
\beq\label{26}
{\wtilde S}_-^{(1)}\ket{s}=0\ , \qquad
{\wtilde S}_0^{(1)}\ket{s}=-s\ket{s}\ , \qquad
\ket{ss_0}=\left({\wtilde S}_+^{(1)}\right)^{s+s_0}\ket{s}\ . 
\eeq
For $\ket{ss_0}$, the relations (\ref{24}) and (\ref{25}) give us 
\beq\label{27}
{\wtilde S}_{0,\pm}^{(2)}\ket{ss_0}=0\ . 
\eeq
The above tells us that the eigenvalue problem for $({\wtilde S}_{0,\pm})$ in our present model 
can be treated in the frame of $({\wtilde S}_{0,\pm}^{(1)})$. 
The part $({\wtilde S}_{0,\pm}^{(1)})$ can be rewritten in the form 
\beq\label{28}
& &{\wtilde S}_0^{(1)}=-s+\frac{1}{2}{\tilde n}\ , \qquad
({\tilde n}=\sum_{\mu}{}^{(1)}({\tilde a}_\mu^*{\tilde a}_\mu +{\tilde b}_{\mu}^*{\tilde b}_\mu)) \nonumber\\
& &{\wtilde S}_+^{(1)}=\sum_\mu{}^{(1)}{\tilde a}_\mu^*{\tilde b}_\mu^*\ , \qquad
{\wtilde S}_-^{(1)}=\sum_\mu{}^{(1)}{\tilde b}_\mu {\tilde a}_\mu\ . 
\eeq
Here, ${\tilde a}_\mu^*$ and ${\tilde b}_\mu^*$ are defined as 
\beq
{\tilde a}_\mu^*={\tilde c}_{1\mu}^*\ , \qquad
{\tilde b}_\mu^*={\tilde c}_{0\mu}\ . \qquad (\mu=1,\ 2,\cdots ,\ 2s)
\nonumber
\eeq
Needless to say, $({\wtilde S}_{0,\pm}^{(1)})$ obeys the $su(2)$-algebra: 
\beq\label{29}
\left[\ {\wtilde S}_+^{(1)}\ , \ {\wtilde S}_-^{(1)}\ \right]=2{\wtilde S}_0^{(1)}\ , \qquad
\left[\ {\wtilde S}_0^{(1)}\ , \ {\wtilde S}_\pm^{(1)}\ \right]=\pm{\wtilde S}_\pm^{(1)}\ .
\eeq
We can see that $({\tilde a}_\mu^*,{\tilde a}_\mu)$ and $({\tilde b}_\mu^*, {\tilde b}_\mu)$ 
are nothing but the particle and the hole operators used in the case of the closed-shell system and if 
$\Omega$ used in the closed-shell system is replaced with $s$, the results are available in the case with $s < \Omega$.

We also decompose $({\wtilde \Lambda}_{0,\pm})$ into the following two parts :
\beq\label{30}
{\wtilde \Lambda}_{0,\pm}={\wtilde \Lambda}_{0,\pm}^{(1)} +{\wtilde \Lambda}_{0,\pm}^{(2)}\ , 
\hspace{9cm}
\eeq
\bsub\label{31}
\beq
& &{\wtilde \Lambda}_0^{(1)}=\frac{1}{2}\sum_{\mu}{}^{(1)}({\tilde c}_{1\mu}^*{\tilde c}_{1\mu}+{\tilde c}_{0\mu}^*{\tilde c}_{0\mu})-s
=\frac{1}{2}\sum_{\mu}{}^{(1)}({\tilde a}_\mu^*{\tilde a}_\mu-{\tilde b}_\mu^*{\tilde b}_\mu) \ , \nonumber\\ 
& &{\wtilde \Lambda}_+^{(1)}=\sum_{\mu}{}^{(1)}{\tilde c}_{1\mu}^*{\tilde c}_{0\mu}^*
=\sum_{\mu}{}^{(1)}{\tilde a}_\mu^*{\tilde b}_\mu\ , 
\quad
{\wtilde \Lambda}_-^{(1)}=\sum_{\mu}{}^{(1)}{\tilde c}_{0\mu}{\tilde c}_{1\mu}
=\sum_\mu{}^{(1)}{\tilde b}_\mu^*{\tilde a}_\mu\ , \quad
\label{31a}\\
& &{\wtilde \Lambda}_0^{(2)}=\frac{1}{2}\sum_{\mu}{}^{(2)}({\tilde c}_{1\mu}^*{\tilde c}_{1\mu}+{\tilde c}_{0\mu}^*{\tilde c}_{0\mu})
-\lambda\ , 
\quad (\lambda=\Omega-s)
\nonumber\\ 
& &{\wtilde \Lambda}_+^{(2)}=\sum_{\mu}{}^{(2)}{\tilde c}_{1\mu}^*{\tilde c}_{0\mu}^*
\ , \quad
{\wtilde \Lambda}_-^{(2)}=\sum_{\mu}{}^{(2)}{\tilde c}_{0\mu}{\tilde c}_{1\mu} \ .
\label{31b}
\eeq
\esub
These decomposed algebras satisfy 
\beq
& &\left[\ {\rm any\ of}\ {\wtilde S}_{0,\pm}^{(1)}\ {\rm and}\ {\wtilde S}_{0,\pm}^{(2)}\ , 
\ {\rm any\ of}\ {\wtilde \Lambda}_{0,\pm}^{(1)}\ {\rm and}\ {\wtilde \Lambda}_{0,\pm}^{(2)}\ \right]=0 \ , 
\label{32}\\
& &{\wtilde \Lambda}_{0,\pm}^{(1)}\ket{s}=0\ , 
\label{33}\\
& & {\wtilde \Lambda}_-^{(2)}\ket{s}=0\ , \qquad
{\wtilde \Lambda}_0^{(2)}\ket{s}=-\lambda\ket{s}\ . \qquad
(\lambda=\Omega-s)
\label{34}
\eeq
Then, by combining the above relations with the form (\ref{20}), we have 
\beq\label{35}
\ket{ss_0,\lambda\lambda_0;s+\lambda=\Omega}
&=&\left({\wtilde \Lambda}_+^{(2)}\right)^{\frac{1}{2}(N-2s)}\ket{ss_0}\nonumber\\
&=&\left({\wtilde \Lambda}_+^{(2)}\right)^{\frac{1}{2}(N-2s)}\left({\wtilde S}_+^{(1)}\right)^{s+s_0}\ket{s}\ . 
\eeq
The relation (\ref{35}) tells us that, in the case where the Hamiltonian (\ref{5}) is investigated, 
at the first step, we specify the single-particle states $\mu=1,\ 2,\cdots,\ 2s$ which give us the 
minimum weight state $\ket{s}$. 
At the next step, we investigate the Hamiltonian (\ref{5}) under appropriate method. 
Then, at the final step, we operate $({\wtilde \Lambda}_+^{(2)})^{\frac{1}{2}(N-2s)}$ on the eigenstates obtained at the 
second step and the eigenstates in the fermion number $N$ are derived. 
The above is our modified form of the $su(2)$-Lipkin model and, in {\bf 3}, 
we formulate the random phase approximation (RPA) 
for the Hamiltonian.

\section{A renewal of the conventional random phase approximation}

In a way similar to that in the conventional random phase approximation (RPA), 
we begin our discussion with the analysis of the following equation of motion: 
\beq\label{36}
\left[\ {\wtilde H}\ , \ {\wtilde S}_{\pm}^{(1)}\ \right]=\pm 2\epsilon {\wtilde S}_{\pm}^{(1)}
\pm 2G\left[\ {\wtilde S}_0^{(1)}\ , \ {\wtilde S}_{\mp}\ \right]_+\ . 
\eeq
The term $\pm 2G[{\wtilde S}_0^{(1)}, {\wtilde S}_\mp]_+$ can be rewritten as 
\beq\label{37}
\pm 2G\left[\ {\wtilde S}_0^{(1)}\ , \ {\wtilde S}_{\mp}\ \right]_+ 
=\mp G\left[\ 2s-{\tilde n}\ , \ {\wtilde S}_{\mp}^{(1)}\ \right]_+
\mp 2G(2s-{\tilde n}){\wtilde S}_{\mp}^{(2)}\ . 
\eeq
In the case with $s=\Omega$, the term $\mp 2G(2s-{\tilde n}){\wtilde S}_{\mp}^{(2)}$ does not exist 
because of ${\wtilde S}_\mp={\wtilde S}_{\mp}^{(1)}$. 
And ${\tilde n}$ regarded as ${\tilde n}=0$, the relation (\ref{36}) is reduced to the starting equation of the 
conventional RPA for the closed shell system, that is, 
$[ {\wtilde H} ,  {\wtilde S}_{\pm}^{(1)} ]$ can be linearized for ${\wtilde S}_{\pm}^{(1)}$. 
In the case with $s < \Omega$, the term  $\mp 2G(2s-{\tilde n}){\wtilde S}_{\mp}^{(2)}$ exists as the 
operator. 
However, as long as the equation of motion is investigated in the orthogonal set given in the 
relation (\ref{26}), this term may be permitted to throw away. 
If we follow the above argument, our problem is to result in developing the idea how to treat ${\tilde n}$. 
Under the assumption that the change of the occupation of the single-particle states is small, 
it might be permitted beforehand to set up ${\tilde n}=0$. 
However, in the case where the interaction strength becomes large, 
the effect coming from ${\tilde n}$ is conjectured to be a subject of our investigation even if $s=\Omega$.

Before coming back to the main subject, we will mention some points which the Hamiltonian (\ref{5}) connotes. 
The Hamiltonian (\ref{5}) is expressed in the linear form for ${\wtilde S}_0$, $({\wtilde S}_+)^2$ and 
$({\wtilde S}_-)^2$ and, then, the eigenstates are classified in terms of 
the linear combinations of the even power of ${\wtilde S}_+^{(1)}$ in the states (\ref{26}). 
Second group consists of the eigenstates expressed in terms of the linear combinations 
of the odd power of ${\wtilde S}_+^{(1)}$ in the states (\ref{26}). 
Former and latter will be denoted by $\rket{2k}$ and $\rket{2k+1}$ $(k=0,1,2,\cdots)$, respectively. 
We regard $\rket{0}$ and $\rket{1}$ $(k=0)$ as the ground- and the first excited-state, respectively. 
Therefore, for the Hamiltonian (\ref{5}), we have 
\beq\label{38}
{\wtilde H}\rket{0}=E_0\rket{0}\ , \qquad
{\wtilde H}\rket{1}=E_1\rket{1}\ , \qquad 
\omega=E_1-E_0\ (\geq 0)\ . 
\eeq
Further, without any approximation, we have the following relations for $k=0,1,2,\cdots$: 
\bsub\label{39}
\beq
& &\rbra{1}{\wtilde S}_{\pm}^{(1)}\rket{2k+1}=\rbra{0}{\wtilde S}_{\pm}^{(1)}\rket{2k}=0\ , 
\label{39a}\\
& &\rbra{1}{\tilde n}\rket{2k}=\rbra{0}{\tilde n}\rket{2k+1}=0 \ . 
\label{39b}
\eeq
\esub
With the aid of the relation $\sum_{k=0}(\rket{2k}\rbra{2k}+\rket{2k+1}\rbra {2k+1})=1$ and 
with the use of the relation (\ref{39}), we have the following :
\bsub\label{40}
\beq
& &\rbra{1}{\tilde n}{\wtilde S}_{\pm}^{(1)}\rket{0}=\sum_{k=0}\rbra{1}{\tilde n}\rket{2k+1}
\rbra{2k+1}{\wtilde S}_{\pm}^{(1)}\rket{0}\ , 
\label{40a}\\
& &\rbra{1}{\wtilde S}_{\pm}^{(1)}{\tilde n}\rket{0}=\sum_{k=0}
\rbra{1}{\wtilde S}_{\pm}^{(1)}\rket{2k}\rbra{2k}{\tilde n}\rket{0}\ . 
\label{40b}
\eeq
\esub
In addition to the relations (\ref{39}), we introduce an approximation for terms related to $\rket{0}$ and $\rket{1}$:
\beq\label{41}
\rbra{2k+1}{\wtilde S}_{\pm}^{(1)}\rket{0}
=\rbra{2k}{\tilde n}\rket{0}=0\quad {\rm for}\quad k\neq 0\ . 
\eeq
Then, the relation (\ref{40}) is approximated as 
\bsub\label{42}
\beq
& &\rbra{1}{\tilde n}{\wtilde S}_{\pm}^{(1)}\rket{0}=\rbra{1}{\tilde n}\rket{1}\rbra{1}{\wtilde S}_{\pm}^{(1)}\rket{0}=n_1S_{\pm}\ , 
\label{42a}\\
& &\rbra{1}{\wtilde S}_{\pm}^{(1)}{\tilde n}\rket{0}=\rbra{1}{\wtilde S}_{\pm}^{(1)}\rket{0}\rbra{0}{\tilde n}\rket{0}=n_0 S_{\pm}\ , 
\label{42b}\qquad\qquad\qquad\qquad
\eeq
\esub
\beq\label{43}
\rbra{1}{\tilde n}\rket{1}=n_1\ , \quad
\rbra{0}{\tilde n}\rket{0}=n_0\ , \quad
\rbra{1}{\wtilde S}_{\pm}^{(1)}\rket{0}=S_{\pm} \ . 
\eeq
The approximate relation (\ref{42}) gives us 
\beq\label{44}
\rbra{1}[\ {\tilde n}\ , \ {\wtilde S}_{\pm}^{(1)}\ ]_+\rket{0}=(n_1+n_0)S_{\pm}\ . 
\eeq

Following the above-mentioned scheme for the approximation, we can derive the following equation from 
the relation (\ref{36}):
\beq\label{45}
\omega S_\pm = \pm 2\epsilon S_{\pm}\mp 2G(2s-n)S_{\mp}\ . 
\eeq
Here, $n$ denotes 
\beq\label{46}
n=\frac{1}{2}(n_1+n_0)=n_0+\frac{1}{2}(n_1-n_0)\ . 
\eeq
In the case with $(s=\Omega,\ n=0)$, we can see that the relation (\ref{46}) is reduced to the 
equation derived under the conventional idea for RPA. 
In the conventional treatment, we have a relation to determine the absolute values of $S_{\pm}$. 
The relation corresponding to the conventional one is derived through the following process: 
With the use of the relation $\rbra{0}[{\wtilde S}_-^{(1)}, {\wtilde S}_+^{(1)}]\rket{0}=\rbra{0}-2{\wtilde S}_0^{(1)}\rket{0}$, 
we have 
\beq\label{47}
\sum_{k=0}\left(\rbra{2k+1}{\wtilde S}_+^{(1)}\rket{0}^2-\rbra{2k+1}{\wtilde S}_-^{(1)}\rket{0}^2\right)
=2s-\rbra{0}{\tilde n}\rket{0}\ . 
\eeq
By adopting the approximation (\ref{41}), the relation (\ref{47}) becomes 
\beq\label{48}
(S_+)^2-(S_-)^2=2s-n_0\ . 
\eeq
In the case with $(s=\Omega, \ n_0=0)$, the relation (\ref{48}) is written as 
\beq\label{49}
\left(S_+/\sqrt{2\Omega}\right)^2-\left(S_-/\sqrt{2\Omega}\right)^2=1\ . 
\eeq
The above is well-known formula under the name of the relation between the forward- and the backward-amplitude. 
The relations (\ref{45}) and (\ref{48}) form the basic equations of the present treatment. 
Needless to say, if $s=\Omega$ and $n_0=0$, both reduce to those in the conventional RPA. 
If $n_0$ and $n_1$ are given, the relations (\ref{45}) and (\ref{48}) give us the following results :
\beq
& &\frac{\omega}{2\epsilon}=\sqrt{1-\left(\frac{G}{\epsilon}(2s-n)\right)^2}\ , 
\quad {\rm i.e.,}\quad 
\frac{G}{\epsilon}(2s-n)=\sqrt{1-\left(\frac{\omega}{2\epsilon}\right)^2}\ , 
\label{50}\\
& &(S_\pm)^2=(2s-n_0)\frac{1}{2}\left(\left(\frac{2\epsilon}{\omega}\right)\pm 1\right)\ . 
\label{51}
\eeq
If $s=\Omega$ and $n_0=n_1=0$, $\omega/(2\epsilon)$ can be expressed as 
\beq\label{52}
\frac{\omega}{2\epsilon}=\sqrt{1-\left(\frac{G}{\epsilon}\cdot 2\Omega\right)^2}\ . 
\eeq
In the case with $G\cdot 2\Omega/\epsilon =1$, $\omega/(2\epsilon)=0$ and at this point, 
the phase transition occurs. 
In our case, also the condition $G\cdot (2s-n)/\epsilon=1$ leads us $\omega/(2\epsilon)=0$. 
However, we can see that the above condition depends on $n$. 
The case $(S_{\pm})^2$ is also in the same situation as the above.

As was mentioned in the above, our further task is to determine $n_0$ and $n_1$ which may be functions of $\omega/(2\epsilon)$. 
Our first task is concerned with $n_0(=\rbra{0}{\tilde n}\rket{0})$. 
The idea comes from the Casimir operator (\ref{4}). 
By calculating the expectation value $\rbra{0}{\wtilde {\mib S}}^2\rket{0}(=s(s+1))$, we have 
\beq\label{53}
(2s-n_0)^2+2\left(\frac{2\epsilon}{\omega}\right)(2s-n_0)-4s(s+1)=0\ . 
\eeq
Here, $\rbra{0}[{\wtilde S}_+^{(1)} , {\wtilde S}_-^{(1)}]_+/2\rket{0}$ and 
$\rbra{0}({\wtilde S}_0^{(1)})^2\rket{0}$ are approximated in the form 
\bsub\label{54}
\beq
& &\rbra{0}\frac{1}{2}[\ {\wtilde S}_+^{(1)}  ,\ {\wtilde S}_-^{(1)}\ ]_+\rket{0}
=\frac{1}{2}\left((S_+)^2+(S_-)^2\right)
=\frac{1}{2}\left(\frac{2\epsilon}{\omega}\right)(2s-n_0)\ , 
\label{54a}\\
& &\rbra{0}({\wtilde S}_0^{(1)})^2\rket{0}=\left(-s+\frac{1}{2}n_0\right)^2=\frac{1}{4}(2s-n_0)^2\ . 
\label{54b}
\eeq
\esub
Since $2s-n_0 \geq 0$, we pick up the positive solution of the above-quadratic equation for $(2s-n_0)$: 
\beq\label{55}
2s-n_0=-\frac{2\epsilon}{\omega}+\sqrt{\left(\frac{2\epsilon}{\omega}\right)^2+4s(s+1)}
=\frac{\omega}{2\epsilon}\cdot \frac{4s(s+1)}{\sqrt{1+4s(s+1)\left(\frac{\omega}{2\epsilon}\right)^2}+1}\ . 
\eeq
The inverse of the above is expressed as 
\setcounter{equation}{54}
\bsub
\beq\label{55a}
\frac{\omega}{2\epsilon}=\frac{1}{\sqrt{4s(s+1)}-(2s-n_0)}-\frac{1}{\sqrt{4s(s+1)}+(2s-n_0)}\ . 
\eeq
\esub
The relation (\ref{55}) will be taken up again in the relations (\ref{72}),  (\ref{73}) and (\ref{74}). 
We can see that the solution (\ref{55}) gives us the following boundary values : 
\bsub\label{56}
\beq
& &{\rm (i)\ If}\ \omega/(2\epsilon)=1\ , \quad 2s-n_0=2s\ , \quad {\rm i.e.,}\quad n_0=0\ , 
\label{56a}\\
& &{\rm (ii)\ If}\ \omega/(2\epsilon)=0\ , \quad 2s-n_0=0\ , \quad {\rm i.e.,}\quad n_0=2s\ . 
\label{56b}
\eeq
\esub
The relation (\ref{55}) can be expressed as a function of $S_-$ : 
\beq\label{57}
n_0=\frac{(S_-)^2}{2s+1+\sqrt{(2s+1)^2-4(S_-)^2}}\ . 
\eeq
The above expression is approximated in the form 
\beq\label{58}
n_0=\frac{(S_-)^2}{2(2s+1)}\ . \quad 
{\rm if}\quad 2s+1\gg 2(S_-)\ , \quad
{\rm i.e.,}\quad 
n_0\ll s+\frac{1}{2}\ \ {\rm or}\ \ n_0\ll S_-\ , 
\eeq
The form (\ref{58}) may be regarded as the simplest approximate expression for $n_0$. 
As a supplementary remark, we mention the following: 
About sixty years ago, the effect of $n_0$ as the quasi-particle number was investigated in the 
study of the quasi-particle RPA for microscopic description of the first excited state of medium 
heavy even-even nuclei\cite{8}. 
If the formula used at that time is translated into the present model, it may be reduced to the form (\ref{58}). 
In this case, $S_-/\sqrt{2s}$ plays a role of the backward amplitude in the quasi-particle RPA. 
Of course, the formula (\ref{58}) is not applicable to the case with $n_0 \approx 2s$.

Next, we discuss our idea for $n_1(=\rbra{1}{\tilde n}\rket{1})$. 
First, we notice the relation (\ref{36}), which can be rewritten as follows : 
\beq\label{59}
\left[\ {\wtilde H}-\epsilon {\tilde n}\ , \ {\wtilde S}_{\pm}\ \right]=\mp G\left[\ 2s-{\tilde n}\ , \ {\wtilde S}_{\mp}\ \right]_+\ . 
\eeq
We discuss the case with $N=2s$ and, then, omit the superscript (1) used in {\bf 2}. 
The relation (\ref{59}) gives us 
\beq\label{60}
\left[\ {\wtilde H}-\epsilon {\tilde n}\ , \ A{\wtilde S}_+ +B{\wtilde S}_-\ \right]
=-\left[\ G(2s-{\tilde n})\ , \ A{\wtilde S}_- -B{\wtilde S}_+\ \right]_+\ . 
\eeq
Associating with the relation (\ref{60}), we take up the following : 
\beq\label{61}
\left[\ {\tilde n}\ , \ A{\wtilde S}_+ +B{\wtilde S}_-\ \right]=2\left(A{\wtilde S}_+ - B{\wtilde S}_-\right) \ . 
\eeq
For arbitrary values of $A$ and $B$, we have the relations (\ref{60}) and (\ref{61}). 
We will show that, for specific values, they are useful for our purpose. 
For the both sides of the relation (\ref{60}), the states $\rket{0}$ and $\rket{1}$ lead us to 
\bsub\label{62}
\beq
& &{\textrm{The\ left-hand\ side}}\nonumber\\
&=&\rbra{1}[\ {\wtilde H}-\epsilon{\tilde n}\ , \ A{\wtilde S}_+ + B{\wtilde S}_-\ ]\rket{0}
=
\left((E_1-\epsilon n_1)-(E_0-\epsilon n_0)\right)(AS_+ +BS_-)\ , 
\label{62a}\\
& &\textrm{The\ right-hand\ side}\nonumber\\
&=&\rbra{1}-[\ G(2s-{\tilde n})\ , \ A{\wtilde S}_- - B{\wtilde S}_+\ ]_+\rket{0}
=
-2G(2s-n)(AS_- -BS_+)\ . 
\label{62b}
\eeq
\esub
Here, $S_{\pm}$ are shown in the relation (\ref{51}) and, for the derivation of the relation (\ref{62}), we adopt the 
approximation (\ref{42}) and $n$ is given in the relation (\ref{46})\ ($n=(n_1+n_0)/2$). 
For the relation (\ref{62b}), we impose the condition 
\beq\label{63}
AS_- -BS_+=0\ .
\eeq
Under this condition, the relation (\ref{62b}) vanishes and $(A, B)$ is of the form 
\beq\label{64}
A=\frac{S_+}{(S_+)^2+(S_-)^2}\cdot C\ , \qquad
B=\frac{S_-}{(S_+)^2+(S_-)^2}\cdot C\ . 
\eeq
Here, $C$ denotes an arbitrary $c$-number factor $(C\neq 0$). 
The term $(AS_+ +BS_-)$ in the relation (\ref{62a}) is expressed as 
\beq\label{65}
AS_+ +BS_- = C \neq 0\ .
\eeq
The above consideration leads us to 
\beq\label{66}
(E_1-\epsilon n_1)-(E_0-\epsilon n_0)=0 \ .
\eeq
Since $E_1-E_0=\omega$, $(n_1-n_0)$ is determined in the form 
\beq\label{67}
n_1-n_0=2\left(\frac{\omega}{2\epsilon}\right)\ . 
\eeq
Combining the relation (\ref{67}) with (\ref{56}) for the boundary values of $n_0$, we have 
\bsub\label{68}
\beq
& &{\rm (i)\ If}\ \ \omega/(2\epsilon)=1\ , \ \ n_0=0\ \ {\rm and,\ then}\ \ n_1=2\ , 
\label{68a}\\
& &{\rm (ii)\ If}\ \ \omega/(2\epsilon)=0\ , \ \  n_0=2s\ \ {\rm and,\ then}\ \ n_1=2s\ . 
\label{68b}
\eeq
\esub

In order to examine the validity of the form (\ref{67}), we apply our idea to the 
relation (\ref{61}). 
With the use of $\rket{0}$ and $\rket{1}$, this relation is expressed in the form 
\beq\label{69}
\rbra{1}[\ {\tilde n}\ , \ A{\wtilde S}_+ +B{\wtilde S}_-\ ]\rket{0}
=2\rbra{1}(A{\wtilde S}_+ - B{\wtilde S}_-)\rket{0}\ . 
\eeq
The above gives us 
\beq\label{70}
(n_1-n_0)(AS_+ + BS_-)=2(AS_+ -BS_-)\ . 
\eeq
With the aid of the form (\ref{64}), we obtain another expression for $(n_1-n_0)$ : 
\beq\label{71}
n_1-n_0=2\cdot \frac{(S_+)^2-(S_-)^2}{(S_+)^2+(S_-)^2}\ .
\eeq
Substituting the relation (\ref{51}) for $(S_{\pm})^2$ into (\ref{71}), we can see that the relation (\ref{71}) is 
reduced to the form (\ref{67}). 
The above argument suggests us that, for special values of $A$ and $B$, the relations (\ref{60}) and (\ref{61}) 
are compatible with each other, even if they are treated under our scheme for the approximation.

\section{Discussion with numerical results}

The $su(2)$-Lipkin model is, needless to say, constructed under the $su(2)$-algebra and 
it contains two basic parameters, $s$ and $G/\epsilon$, the numerical values of which are given from the outside. 
The parameter $\epsilon$ plays only a role of the measure of the energy. 
Therefore, the quantity such as, for example, $\omega/(2\epsilon)$, should be specified in terms of a function of 
$s$ and $G/\epsilon$. 
However, as can be seen in the relation (\ref{50}) together with (\ref{55}) and (\ref{67}), 
$G/\epsilon$ is simply expressed as a function of $\omega/(2\epsilon)$. 
But, the inverse seems to be too complicated to give the explicit form. 
In this section, mainly, we will try to search the explicit form 
by incorporating the numerical results in the discussion.

Let us start with the discussion on $n_0$ and $n_1$. 
As was already mentioned, $G/\epsilon$ can be easily expressed as a function of $\omega/(2\epsilon)$. 
We can express $(2s-n)$ in the form 
\beq\label{72}
2s-n&=&
2s-n_0-\frac{1}{2}(n_1-n_0)\nonumber\\
&=&\frac{\omega}{2\epsilon}\left[\frac{4s(s+1)}{\sqrt{1+4s(s+1)\left(\frac{\omega}{2\epsilon}\right)^2}+1}-1\right]\ . 
\qquad \left(\frac{1}{2}(n_1-n_0)=\frac{\omega}{2\epsilon}\right)
\eeq
The quantity $\omega/(2\epsilon)$ changes its value in the range $0\leq \omega/(2\epsilon) \leq 1$ and 
we notice the following inequality :
\beq\label{73}
{\rm if}\ \ s \gg 0\ , \qquad 
\frac{4s(s+1)}{\sqrt{1+4s(s+1)\left(\frac{\omega}{2\epsilon}\right)^2}+1}\gg 1\ . 
\eeq
For example, the case with $s=5$ and $\omega/(2\epsilon)=1/2$ gives us 
$120/(\sqrt{31}+1)=18.271 \gg 1$. 
This indicates that the first term in the bracket $[\quad ]$ in the relation (\ref{72}) is much larger than the second, 1. 
Then, it may be permitted to neglect the second term, 1 and as a possible form of $(2s-n)$, we can adopt the form 
\beq\label{74}
2s-n=L
\sqrt{4s(s+1)}
\ . 
\eeq
Here, $L$ is defined in the form 
\beq\label{75}
L=\frac{
\sqrt{4s(s+1)\left(\frac{\omega}{2\epsilon}\right)^2}}
{\sqrt{1+4s(s+1)\left(\frac{\omega}{2\epsilon}\right)^2}+1}\  .
\eeq
The above approximation corresponds to the condition $n_1=n_0$.

By substituting the relation (\ref{74}) into (\ref{50}), we obtain $G/\epsilon$ which is expressed in terms of $\omega/(2\epsilon)$ :
\beq\label{76}
G/\epsilon\cdot \sqrt{4s(s+1)}L=\sqrt{1-(\omega/(2\epsilon))^2}\ . 
\eeq
The relation (\ref{76}) gives us $\omega/(2\epsilon)$ expressed in terms of $G/\epsilon$ and $s$. 
If $s$ is finite, it may be easy to see that relation (\ref{76}) leads us to the following two cases :
\beq\label{77}
{\rm (i)\ if}\ \ \omega/(2\epsilon)=1\ , \quad G/\epsilon=0\ , \quad& & 
{\rm if}\ \ G/\epsilon=0\ , \quad \omega/(2\epsilon)=1\ , \nonumber\\
&{\rm inversely,}\nonumber\\
{\rm (ii)\ if}\ \ \omega/(2\epsilon)\rightarrow 0\ , \quad G/\epsilon\rightarrow \infty\ , & & 
{\rm if}\ \ G/\epsilon\rightarrow \infty\ , \quad \omega/(2\epsilon)\rightarrow 0\ .
\eeq
In the conventional RPA, the case (i) is familiar to us, but, the case (ii) is ununderstandable. 
Especially, the case (ii) informs us that we cannot expect the phase transition which is well known in the conventional RPA. 
Therefore, it may be interesting to investigate the present RPA in parallel 
with the conventional RPA. 
For this aim, we introduce an auxiliary parameter $g/\epsilon$ which plays a role similar to that of $G/\epsilon$:
\beq\label{78}
g/\epsilon\cdot 2s=G/\epsilon\cdot \sqrt{4s(s+1)}L\ . 
\eeq
The relation (\ref{76}) can be expressed in the following form by introducing the parameter $\delta$ : 
\beq\label{79}
\delta=g/\epsilon\cdot 2s=\sqrt{1-(\omega/(2\epsilon))^2}\ , \qquad
{\rm i.e.,}\qquad
\omega/(2\epsilon)=\sqrt{1-\delta^2}\ . 
\eeq
Formally, the relation (\ref{79}) appears in the conventional RPA described under the interaction strength $g/\epsilon$. 
Associating with $\delta$, we define $\delta_0$ in the form 
\beq\label{80}
\delta_0=G/\epsilon\cdot \sqrt{4s(s+1)}\ . 
\eeq
The relations (\ref{75}) and (\ref{78})$\sim$(\ref{80}) give us to the following relation: 
\beq\label{81}
(\delta_0^2-\delta^2)\sqrt{1-\delta^2}=\left(\frac{\delta_0}{\sigma}\right)\delta\ . 
\qquad (\sigma=\sqrt{4s(s+1)}/2
)
\eeq
With the aid of the relation (\ref{81}), $\delta$ can be expressed as a function of $\delta_0$ and $s$. 
Of course, $\delta_0$ and $\delta$ are restricted to the condition
\bsub\label{82}
\beq
& &{\rm (A)}\ \ 0\leq \delta_0 \leq 1\ , \qquad 0\leq \delta \leq \delta_0\ , 
\label{82a}\\
& &{\rm (B)}\ \ 1\leq \delta_0\ , \qquad \qquad 
0\leq \delta \leq 1\ . 
\label{82b}
\eeq
\esub
Squaring both sides of the relation (\ref{81}), we have
\beq\label{83}
(\delta_0^2-\delta^2)^2(1-\delta^2)=\left(\frac{\delta_0}{\sigma}\right)^2\delta^2\ . 
\eeq
We can see that the relation (\ref{83}) is a cubic equation for $\delta^2$. 
%
%
For example, the case with $\delta_0=1$ is easily solved in the exact form
\beq\label{84}
\delta=\sqrt{1-\frac{1}{\sqrt[3]{2\sigma^2}}\left(\sqrt[3]{\sqrt{1+\frac{4}{27\sigma^2}}+1}
-\sqrt[3]{\sqrt{1+\frac{4}{27\sigma^2}}-1}\right)}
\ (=\delta^1)\ .
\eeq
Of course, the above obeys the condition (\ref{82}). 
However, the process for arriving at the exact solutions in more general cases is too troublesome 
to treat directly the cubic equation (\ref{83}). 
Then, we will try to find a possible approximate solutions. 
For this task, we use the solution (\ref{84}).

Under the above preparative consideration, we will search an approximate method. 
The relation (\ref{81}) is regarded as a quadratic equation for $\delta_0$ and the solution obeying the condition (\ref{82}) 
is given by 
\beq\label{85}
\delta_0=\frac{\delta}{2\sigma\sqrt{1-\delta^2}}\left(1+\sqrt{1+\left(2\sigma\sqrt{1-\delta^2}\right)^2}\right)\ . 
\eeq
Behavior of the relation (\ref{85}) is depicted in Fig.3 for the case $s=5$. 
%
\begin{figure}[t]
\begin{center}
\includegraphics[height=6.0cm]{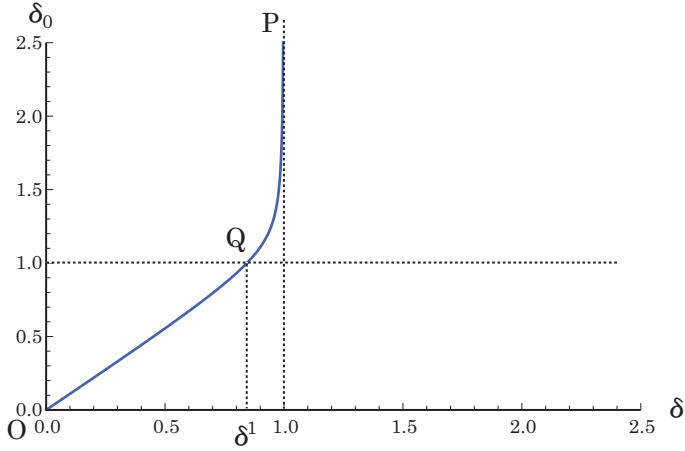}
\caption{The exact solutions $\delta_0$ in Eq.(\ref{85}) are depicted as a function of $\delta$. 
}
\label{fig:3}
\end{center}
\end{figure}
%
The curve starts from the point O ($\delta=0,\ \delta_0=0$) under the derivative 
$(d\delta_0/d\delta)_{\delta=0}=\sqrt{(s+1)/s}$ and, via the point Q ($\delta=\delta^1,\ \delta_0=1$) 
with the derivative $(d\delta_0/d\delta)_{\delta=\delta^1}=(2(\delta^1)^2+1)/(3\delta^1-1)$, 
it gets closed to the point P ($\delta\rightarrow 1,\ \delta_0\rightarrow \infty$). 

%
\begin{figure}[b]
\begin{center}
\includegraphics[height=6.0cm]{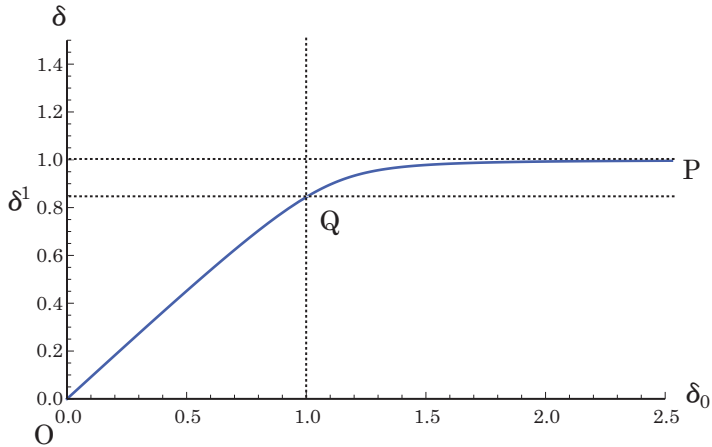}
\caption{The exact solutions $\delta$ in Eq.(\ref{85}) are depicted as a function of $\delta_0$. 
}
\label{fig:4}
\end{center}
\end{figure}
%

Then, our problem is reduced to depict $\delta$ as a function of $\delta_0$ and 
it is carried out by exchanging the axis $\delta_0$ for the axis $\delta$ in Fig.3. 
In this way, we get Fig.4. 
For this task, the derivatives $(d\delta/d\delta_0)_{\delta_0=0}=\sqrt{s/(s+1)}$ and 
$(d\delta/d\delta_0)_{\delta_0=1}=(3\delta^1-1)/(2(\delta^1)^2+1)$ are used and the 
others do not change. 
For an approximate expression of the curve 
OQP in Fig.4, we set up the form : 
\beq\label{86}
\delta=\frac{F(\delta_0)}{\sqrt{1+F(\delta_0)^2}}\ . 
\eeq
The reason why we set up the from (\ref{86}) can be given as follows : 
Figure 4 suggests us that the curve OQP resembles the function $\sin \theta$ in the 
range $0\leq \theta \leq \pi$. 
Of course, $\theta$ may be very complicated function of $\delta_0$. 
Then, let us image a right triangle $\triangle{ABC}$ shown in Fig.5. 
%
\begin{figure}[t]
\begin{center}
\includegraphics[height=6.0cm]{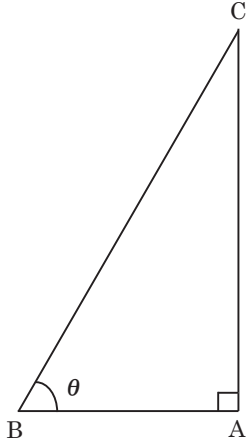}
\caption{A right triangle $\triangle{ABC}$ 
}
\label{fig:5}
\end{center}
\end{figure}
%
In the case where the lengths of three sides of $\triangle{ABC}$ are $AB=1$, $AC=F(\delta_0)$ and 
$BC=\sqrt{1+F(\delta_0)^2}$, we have $\sin \theta =F(\delta_0)/\sqrt{1+F(\delta_0)^2}$. 
Then, following the resemblance mentioned above, it may be permitted to set up 
the relation (\ref{86}). 
The above is the reason why we set up the form (\ref{86}). 
For the relation (\ref{86}), we have 
\beq\label{87}
F(\delta_0)=\frac{\delta}{\sqrt{1-\delta^2}}\ , \qquad
\frac{dF(\delta_0)}{d\delta_0}=\frac{\frac{d\delta}{d\delta_0}}{(\sqrt{1-\delta^2})^3}\ . 
\eeq

As for $F(\delta_0)$, we adopt the following form : 
\beq\label{88}
F(\delta_0)=a_0+a_1\delta_0+a_2(\delta_0)^2+a_3(\delta_0)^3\ . 
\eeq
Here, $a_0$, $a_1$, $a_2$ and $a_3$ denote the constants determined under the 
condition that $\delta$ passes the points O, Q and P with the derivatives at the points O and Q already shown. 
The result is as follows : 
\beq\label{89}
& &a_0=0\ , \qquad a_1=\sqrt{\frac{s}{s+1}}\ , \nonumber\\
& &a_2=\frac{3\delta^1}{\sqrt{1-(\delta^1)^2}}-\frac{1}{(\sqrt{1-(\delta^1)^2})^3}\cdot \frac{3\delta^1-1}{2(\delta^1)^2+1}
-2\sqrt{\frac{s}{s+1}}\ , \nonumber\\
& &a_3=\frac{1}{(\sqrt{1-(\delta^1)^2})^3}\cdot \frac{3\delta^1-1}{2(\delta^1)^2+1}-\frac{2\delta^1}{\sqrt{1-(\delta^1)^2}}
+\sqrt{\frac{s}{s+1}}\ . 
\eeq
For the comparison, the approximate curve is shown together with the exact one in Fig.6. 
The agreement is quite good.

%
\begin{figure}[t]
\begin{center}
\includegraphics[height=6.0cm]{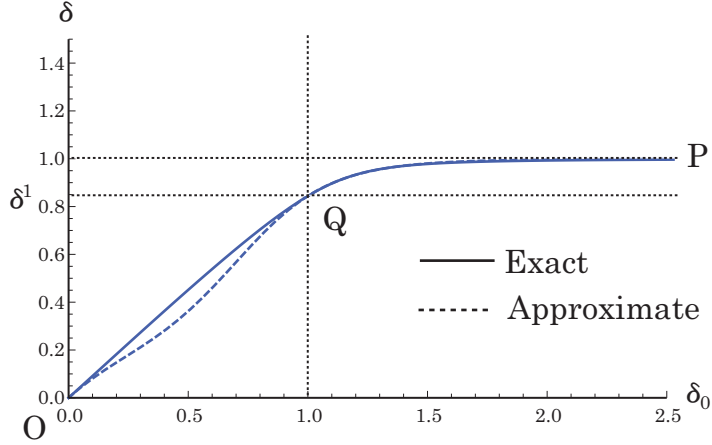}
\caption{The comparison of the exact solutions $\delta$ in Eq.(\ref{85}) and the approximate expression in Eq.(\ref{88}) is 
depicted as a function of $\delta_0$. 
}
\label{fig:6}
\end{center}
\end{figure}
%


In Fig.1, the exact and the RPA results for the first-excited energy in the $su(2)$-Lipkin model are depicted. 
It is ready to compare our result developed in this paper with the exact and the RPA results. 
In Fig.7, the first-excited energy from the ground-state energy is given in Eq.(\ref{79}) with Eq.(\ref{78}) with $s=5$ 
(dash-dotted curve) together with the RPA (dotted curve) and the exact (solid curve) results 
as functions of ${\overline G}(=G/\epsilon\cdot 2\Omega)$ which is related to $\delta_0$ through Eq.(\ref{80}). 
For small $\overline{G}$, our results are close to the RPA results. 
However, in the region before and after $\overline{G}\approx 1$, 
our approximate results go away from the RPA results and 
approximate curve draws closer to the exact results asymptotically for $\overline{G}\gg 1$.  
 
%
\begin{figure}[b]
\begin{center}
\includegraphics[height=6.5cm]{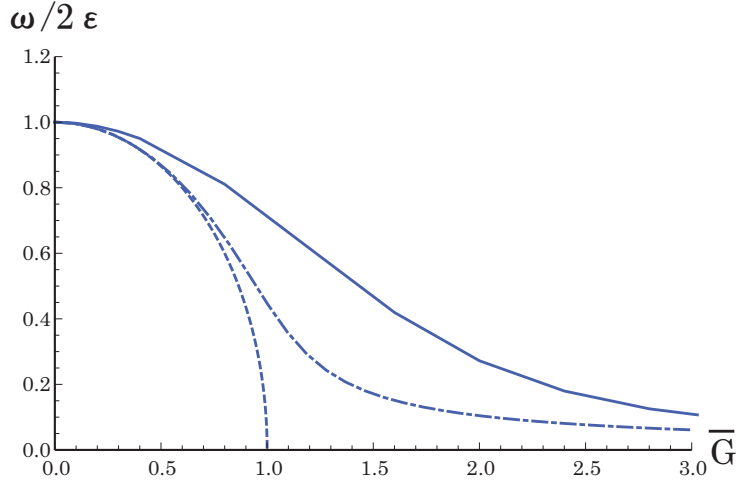}
\caption{The first excited energies from the ground-state energy are depicted as functions 
of $\overline{G}=G/\epsilon\cdot 2\Omega$ for exact (solid curve), the RPA (dashed curve) results with $\Omega=5$ 
and our new 
approximate result (dash-dotted curve) in Eqs.(\ref{79}) and (\ref{78}) with $s=5$. 
}
\label{fig:7}
\end{center}
\end{figure}
%

%
\begin{figure}t]
\begin{center}
\includegraphics[height=6.0cm]{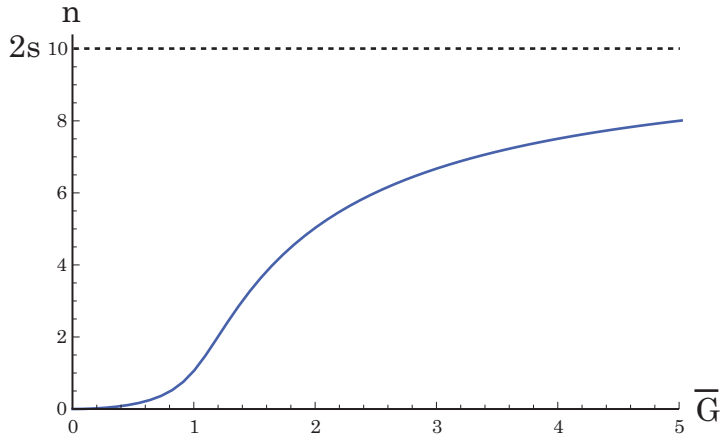}
\caption{In the renewed RPA, $n$ is depicted as a function of ${\overline G}$ with $s=5$.
}
\label{fig:3}
\end{center}
\end{figure}
%

Here, it is interesting to see the value $n\equiv (n_1+n_0)/2$ introduced in {\bf 3}. 
As was already mentioned, if $n_0=n_1=0$, namely $n=0$, then Eq. (\ref{52}) was deduced with $s=\Omega$. 
This result is nothing but the one derived in the conventional RPA as is depicted in Fig.7. 
However, in our renewed RPA, $n_0$ and $n_1$ are determined within the framework developed in {\bf 3}, and 
as a result, $n$ is obtained. 
In Fig.8, the behavior of $n$ is shown numerically as a function of ${\overline G}$.   
For small $\overline{G}$ closer to zero, $n$ is nearly equal to zero. 
Therefore, the conventional RPA results are almost obtained as is shown in Fig.7.
However, in the region before and after $\overline{G}\approx 1$, 
$n$ moves apart from zero. 
Thus, it may be good to say that our approximate results for the first-excited energy in Fig.7 
go away from the conventional RPA results and 
our renewed RPA results draw closer to the exact ones for $\overline{G}\gg 1$.

In conclusion, it would be said that the renewed random phase approximation developed in this paper is 
workable beyond the conventional RPA in all the region of the interaction strength $G$ in the $su(2)$-Lipkin model.

\section{Concluding remarks}

In this paper, we have discussed two problems, which connote the $su(2)$-Lipkin model. 
Through the discussion on the first, we gave a relation among the three quantities $2\Omega$, $N$ and $s$. 
The result is depicted in Fig.2. 
Conventionally, the $su(2)$-Lipkin model has been applied only to the case with $2\Omega=N=2s$, i.e., 
the closed shell nuclei. 
However, our idea supports that we can treat this model also in the case satisfying the inequality (\ref{19}). 
In the second, preserving the framework of the conventional RPA, we developed an idea how to investigate the term 
neglected in the conventional treatment. 
The result is depicted in Fig.7, which suggests us that we must reconsider the terminology, 
the ``phase" used in various situations. 
For example, the two-level pairing model involves two phases, 
the non-superconductive and the superconductive phase in RPA. 
Figure 1 suggests us the following : 
\break
The phase, which is usually interested in and may be called the normal phase, is effective in the range 
$0\leq G/(2\epsilon)\cdot 2s \leq 1$. 
In the range $1< G/(2\epsilon)\cdot 2s$, we cannot see anything. 
This is the result based on the conventional RPA. 
On the other hand, our RPA gives us the 
frequency $\omega/(2\epsilon)\ (>0)$ in any range of $G/(2\epsilon)\cdot 2s$ and, moreover, 
the result approaches the exact one. 
In this sense, we can treat the $su(2)$-Lipkin model only in the frame of the normal phase. 
The above is the second problem. 
If we take these two conclusions into consideration, it may be necessary to investigate another phase, 
which may be called the deformed phase.

With the aid of the Caylee-Klein parameters, the deformed phase in the $su(2)$-Lipkin model may be formulated. 
A new mean field may be constructed by the following canonical transformation from the original fermions 
$({\tilde c}_{1\mu},\ {\tilde c}_{0\mu}^*)$ to the new ones $({\tilde d}_{1\mu},\ {\tilde d}_{0\mu}^*)$ for 
$\mu=1,\ 2,\cdots ,\ 2\Omega$ : 
\beq\label{5-1}
\left(
\begin{array}{c}
{\tilde d}_{1\mu}^* \\
{\tilde d}_{0\mu}^*
\end{array}
\right)
=
\left(
\begin{array}{cc}
u & -v \\
v^* & u^*
\end{array}
\right)
\left(
\begin{array}{c}
{\tilde c}_{1\mu}^* \\
{\tilde c}_{0\mu}^*
\end{array}
\right)
\ , \quad 
{\rm i.e.,}
\quad
\left(
\begin{array}{c}
{\tilde c}_{1\mu}^* \\
{\tilde c}_{0\mu}^*
\end{array}
\right)
=
\left(
\begin{array}{cc}
u^* & v \\
-v^* & u
\end{array}
\right)
\left(
\begin{array}{c}
{\tilde d}_{1\mu}^* \\
{\tilde d}_{0\mu}^*
\end{array}
\right) \ . 
\eeq
Here, $(u,\ v)$ and their complex conjugate are related to the Caylee-Klein parameters, 
the set of which is well known in the dynamics of rigid body. 
They satisfy the relation 
\beq\label{5-2}
|u|^2+|v|^2=1\ .
\eeq
The set $(u,\ v)$ is related with the $D$-function with $j=1/2$ and in terms of the Euler angles $(\phi, \ \theta,\ \psi)$, 
they can be expressed as 
\beq\label{5-3}
u=e^{i\frac{1}{2}(\phi+\psi)}\cos\frac{\theta}{2}\ , \qquad
v=e^{i\frac{1}{2}(\phi-\psi)}\sin \frac{\theta}{2}\ . 
\eeq
The transformation (\ref{5-1}) does not violate the basic framework of the model : 
After transformation, still the model consists of two single-particle levels with the degeneracy $2\Omega$ 
and, further, the forms of the total fermion number operator, the Casimir operator and ${\wtilde \Lambda}_{\pm,0}$ are unchanged. 
Of course, the Hamiltonian changes its form. 
Therefore, it may be an interesting problem to investigate the deformed phase. 
Finally, we stress that the $su(2)$-Lipkin model still connotes the problem to give the solution.

\section*{Acknowledgment}

Two of the authors (Y.T. and M.Y.) would like to express their thanks to Professor\break
J. da Provid\^encia and Professor 
C. Provid\^encia, two of the co-authors of this paper for their warm hospitability during their visit to 
Coimbra in the spring of 2015. 
At that time, the research connected with the problem discussed in this paper was started through the 
private meeting by the present four authors. 
Unfortunately, all the four members will be not able to have again such a meeting as the above. 
\let\doi\relax

\let\doi\relax


\end{document}